\def\year{2021}\relax
\documentclass[letterpaper]{article} 
\usepackage{aaai21}  
\usepackage{times}  
\usepackage{helvet} 
\usepackage{courier}  
\usepackage[hyphens]{url}  
\usepackage{graphicx} 
\usepackage{natbib}  
\usepackage{caption} 
\usepackage[switch]{lineno}
\usepackage{kotex}
\usepackage{multirow}
\usepackage{subcaption}     
\usepackage{hyperref}       
\usepackage{amsmath, amssymb}        
\usepackage[capitalise]{cleveref}       
\usepackage{graphicx}
\usepackage{sidecap}        
\sidecaptionvpos{figure}{c}
\newcommand{\etal}{\textit{et al.}}
\usepackage{booktabs}
\usepackage{xcolor}

\title{End-to-End Differentiable Learning to HDR Image Synthesis for Multi-exposure Images}
\author{
    Jung Hee Kim\textsuperscript{\rm *,1}, Siyeong Lee\textsuperscript{\rm *,2}, Suk-Ju Kang\textsuperscript{\rm 1}
    \\
}
\affiliations{
    \textsuperscript{\rm 1} Department of Electronic Engineering Sogang University, Seoul, Korea\\
    \textsuperscript{\rm 2} NAVER LABS, Bundang, Korea\\
    \textsuperscript{\rm *} Equal contribution\\
    kjhe129@sogang.ac.kr, siyeong.lee@naverlabs.com, sjkang@sogang.ac.kr
}

\begin{document}
\maketitle
\begin{abstract}
Recently, high dynamic range (HDR) image reconstruction based on the multiple exposure stack from a given single exposure utilizes a deep learning framework to generate high-quality HDR images. These conventional networks focus on the exposure transfer task to reconstruct the multi-exposure stack. Therefore, they often fail to fuse the multi-exposure stack into a perceptually pleasant HDR image as the inversion artifacts occur. We tackle the problem in stack reconstruction-based methods by proposing a novel framework with a fully differentiable high dynamic range imaging (HDRI) process. By explicitly using the loss, which compares the network's output with the ground truth HDR image, our framework enables a neural network that generates the multiple exposure stack for HDRI to train stably. In other words, our differentiable HDR synthesis layer helps the deep neural network to train to create multi-exposure stacks while reflecting the precise correlations between multi-exposure images in the HDRI process. In addition, our network uses the image decomposition and the recursive process to facilitate the exposure transfer task and to adaptively respond to recursion frequency. The experimental results show that the proposed network outperforms the state-of-the-art quantitative and qualitative results in terms of both the exposure transfer tasks and the whole HDRI process.
\end{abstract}

\section{Introduction}

\noindent Recently, various applications use high dynamic range imaging (HDRI) technique because it provides better aesthetic appreciation than ordinary imaging techniques with a limited dynamic range \cite{sen2016practical}. Moreover, HDRI aims to restore under-exposed and over-exposed regions, so that the reconstructed high dynamic range (HDR) images convey much information such as image details, irrespective of the illuminance change. Especially, recent vision systems have used HDRI to improve their performance in terms of robustness and consistency (e.g., passing through the tunnel). In this context, many approaches such as fusing the multi-exposure stack \cite{debevec2008recovering}, implementing the event cameras \cite{wang2019event} have been introduced to generate high-quality images with gamings \cite{khaldieh2018tone} and sports \cite{weber20154k}. 

Deep neural networks, especially convolutional neural networks (CNNs), have shown their significant role in reconstructing the HDR image. Two primary approaches exist in reconstructing the HDR image: direct reconstruction methods \cite{eilertsen2017hdr,marnerides2018expandnet, liu2020single} and multi-exposure stack-based synthesis methods \cite{endo2017deep, lee2018deepa, lee2018deepb}. Direct reconstruction aims to recover a HDR image (32bits/pixel) from a given single low dynamic range (LDR) image (8bits/pixel). In this case, a large number of LDR-HDR image pair data is required to train a deep neural network \cite{endo2017deep}. There have been many attempts to solve the data quantity problem by crawling image pairs from the internet \cite{kim2019deep} or generating synthetic image pairs \cite{liu2020single}. On the other hand, HDR synthesis with the multi-exposure stack focuses on transferring exposures to generate the multi-exposure stack accurately. These approaches alleviate the dataset quantity problem as they require much fewer scenes with multi-exposure stack \cite{lee2018deepa, lee2018deepb}. However, they suffer from severe local inversion artifacts due to the limitations of networks being trained only with the ground truth multi-exposure stack's supervision. Therefore, the conventional approaches had difficulties training the network in an end-to-end manner to reflect the whole HDRI process.

We propose the differentiable HDR synthesis process, which enables the end-to-end training procedure and alleviates the generation of the local inversion artifacts. We also incorporate the image decomposition approach to disentangle an exposure transfer task and the recurrent network to gradually increase or decrease the exposure level to reconstruct a multi-exposure stack from a single exposure image. In summary, our contributions are three-fold as follows:

\begin{itemize}
\item We propose a novel framework with a differentiable HDRI synthesis method. To overcome the conventional limitations of multi-exposure stack-based HDR synthesis, we applied the differentiable CRF function, which converts discrete pixel intensity values into luminance values in the standard HDRI. By back-propagating the gradient of the loss between the network's outputs and ground truth HDR images explicitly, the networks can escape the local optimum which only focuses on the exposure transfer task, so that generates high-quality HDR images without the local inversion artifacts.

\item We incorporate the image decomposition method for reconstructing the HDR image to focus on preserving the image details in exposure transfer tasks. We disentangle exposure transfer tasks with the two-pathway approach, which adjusts the global tone and reconstructs the local structure of the image individually.

\item We propose a recurrent approach in the multi-exposure stack generation to efficiently utilize the recursive process. Our network learns to generate sequential images with multiple exposures in the recurrent structure as the recursive process requires to maintain gradients until the entire multi-exposure stack is generated.
\end{itemize}

\section{Related Works}
\subsection{Radiometric calibration} Recovering the scene luminance with given LDR images and reconstructing HDR images requires estimating the intensity-to-luminance mapping function of the individual camera. The estimating process of the mapping function is called the radiometric calibration. The commonly used radiometric calibration estimates mapping function, which is the camera response function (CRF), from a given multi-exposure stack and corresponding exposure values. Based on the assumption about the shape of the CRF, most approaches can be categorized into two classes: parametric and non-parametric methods. 

The parametric methods assume the CRF to have a specific and analytic functional form, such as a gamma function\cite{mann1994beingundigital}, or a polynomial function \cite{mitsunaga1999radiometric}. Furthermore, Grossberg and Nayar \cite{grossberg2003determining} modeled CRF using the principal component analysis (PCA) to collect vectors from a large number of real CRFs. Besides, PCA based modeling methods were incorporated into recent deep learning methods \cite{li2017crf, liu2020single}. However, parametric approaches suffer from making explicit assumptions on the analytical form of the CRFs, which is not adequate for monotonic modern camera configurations \cite{chen2019analyzing}.

Non-parametric methods focus on estimating the CRF in a discrete function with the lookup table structure. Debevec and Malik \cite{debevec2008recovering} proposed a least-square formulation with the smoothness constraints to recover CRF in discrete function form. Lee \etal \cite{lee2012radiometric} utilized the observation that images in the multi-exposure stack are linearly dependent on reconstructing HDR images. Badki \etal \cite{badki2015robust} proposed a radiometric calibration method to compensate significant motions in images using a random sample consensus (RANSAC)-based method. Furthermore, a recent deep learning-based approach \cite{endo2017deep, lee2018deepb} applied the most commonly used non-parametric method: Debevec and Malik's approach recovering the CRF. However, since the non-parametric radiometric calibration recovers a CRF as a discrete function and non-differentiable form, the whole end-to-end network implementation considering the multi-exposure stack has been limited.

\subsection {Deep learning-based HDR reconstruction} 
\subsubsection{Direct HDR reconstruction} The recent development of deep neural networks has imposed on learning the direct mapping function between a single LDR image and a target HDR image. Direct methods generate the HDR image without fusing the image stack of different exposures, thereby removing the ghosting artifacts because a spatially aligned multi-exposure image stack is not required. Eilerstsen \etal \cite{eilertsen2017hdr} focused on restoring saturated regions of the under-exposed LDR image to recover the luminance map, which is combined with the input LDR image to reconstruct the HDR image. Marnerides \etal \cite{marnerides2018expandnet} proposed a CNN model that trains to infer a direct mapping function between LDR and HDR images. Khan \etal \cite{khan2019fhdr} implemented feedback structure in reconstructing the HDR image by iteratively refining the HDR reconstruction result. To overcome the dataset quantity challenge, Kim \etal \cite{kim2019deep} and Liu \etal \cite{liu2020single} utilized the dynamic range constrained dataset, which consists of images crawled and extracted from the Internet, and the virtual dataset, respectively. However, since the datasets have diverse dynamic ranges, the normalization process and standardization process for the images become difficult. Due to the undetermined dynamic range of images, the models might be trained in the wrong direction, on account of the gap between virtually generated images and real images. 

\subsubsection{Multi-exposure stack HDR synthesis} Multi-exposure stack HDR synthesis methods incorporated the deep neural network to generate a multi-exposure image stack. The dataset quantity problem of the direct methods can be compensated with a multi-exposure stack method as an arbitrary number of images with different exposures can be used as the training set. The ambiguity of the LDR-to-HDR mapping relation is avoided by focusing on the intermediate task of generating a multi-exposure stack. Endo \etal \cite{endo2017deep} and Lee \etal \cite{lee2018deepa, lee2018deepb} focused on reconstructing the multi-exposure stack from a single LDR image to synthesize a target HDR image. However, these approaches caused the generation of severe local inversion artifacts on reconstructed HDR image as the methods were not trained in the end-to-end structure, where the pixel-wise relations were not imposed. 

\begin{figure*}[ht]
  \centering
  \includegraphics[width=0.68\textwidth]{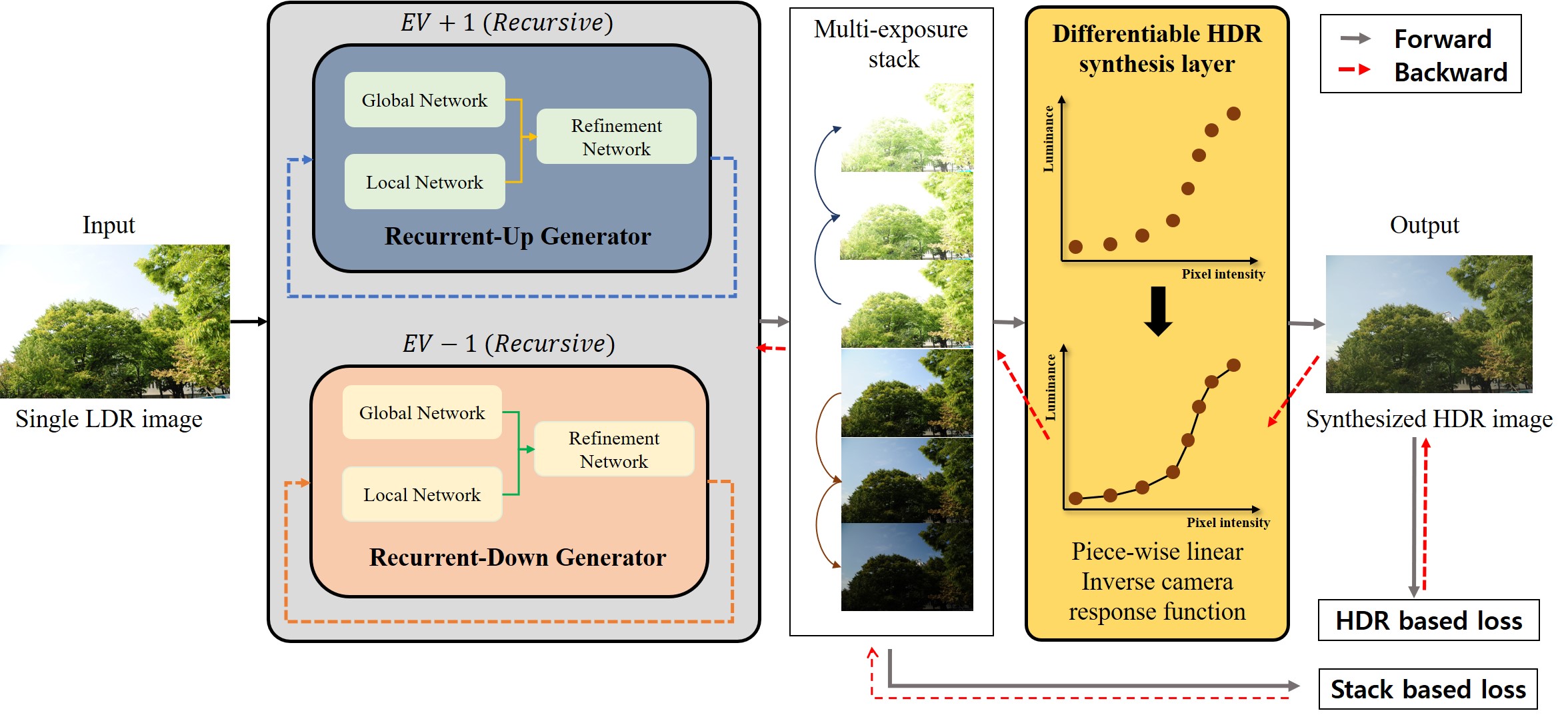}
  \caption{The overall structure of the proposed framework. Our model consists of recurrent-up and recurrent-down networks with the differentiable HDR synthesis layer. Given an input LDR image, the multi-exposure image stack is generated with recursions. Then, the generated stack is synthesized to reconstruct the HDR image with the estimated camera response function using \cref{eq1}.}\label{fig_model}
\end{figure*}

\section{End-to-End Differentiable Learning to HDR Image Synthesis}\label{headings}
This section describes our end-to-end differentiable learning framework that trains both the exposure transfer process for multi-exposure stack generation and the HDR image synthesis, as shown in \cref{fig_model}. We first generate the multi-exposure stack with the recursive process for recurrent-up and recurrent-down networks to reconstruct the entire stack. We then synthesize the stack with the differentiable HDR synthesis layer to reconstruct the HDR image and train the network in the end-to-end structure. We also describe our recurrent network that restores details in saturated regions of the multi-exposure stack by incorporating the image decomposition approach. 
\subsection{Differentiable HDR synthesis layer}
Debevec and Malik \cite{debevec2008recovering} proposed the HDRI pipeline that estimates the CRF using the non-parametric radiometric calibration, which is commonly used. Given LDR images with different exposures, estimating the CRF or inverse CRF is modeled as the least-square problem as follows:
\begin{equation}\label{eq1}
\begin{split}
&O = \sum_{i}^{N}\sum_{j}^{P} [g(Z_{ij}) - \ln E_i + EV_j]^2 \\
&+ \lambda\sum_{z=Z_{min}+1}^{Z_{max}-1}g''(z)^2,
\end{split}
\end{equation}
where $O$ denotes an objective function, $g$ denotes an inverse CRF, and $Z_{ij}$ as a pixel intensity value of $i$-th pixel with $j$-th exposure value. $Z_{min}$ and $Z_{max}$ indicate minimum and maximum intensity values of given LDR images. $N$ and $P$ are the number of images and exposure values of the stack, and $i$, and $j$ are their corresponding indices, respectively. $E_i$ denotes the luminance value of $i$-th pixel and $EV_j$ denotes the $j$-th exposure value. The exposure value can substitute exposure time with a fixed aperture and ISO value. The second term of the objective function regularizes the CRF to be smoothened with the hyperparameter $\lambda$. By minimizing the objective function, we can obtain the discrete CRF of $g$, which maps 8-bit pixel intensity values to 32-bit luminance values. With the recovered inverse CRF $g$, the pixel intensity value can be remapped to the luminance value as follows:
\begin{equation}\label{eq2}
    \ln{E_i}=\ g(Z_{ij})-EV_j.
\end{equation}

The scene luminance is remapped with \cref{eq2}; however, as inverse CRF has the form of the non-differentiable function, we transform the inverse CRF with a linear approximation technique.

Let an inverse CRF be $g=[p_0, p_1, \cdots, p_N]$ with $N$ denoting the maximum intensity value of multi-exposure images. We define the derivative of the linearized function $\hat{g}$ as follows:
\begin{equation}\label{eq3}
\frac{\partial \hat{g}}{\partial Z_{ij}} =  \left\{
    \begin{array}{lr}
        g(0), & \mbox{if } Z_{ij} = 0 \\
        g(Z_{ij}) - g(Z_{ij}-1), &\mbox{otherwise.} 
    \end{array}
\right.
\end{equation}

\begin{figure*}[t]
  \centering
  \includegraphics[width=0.67\textwidth]{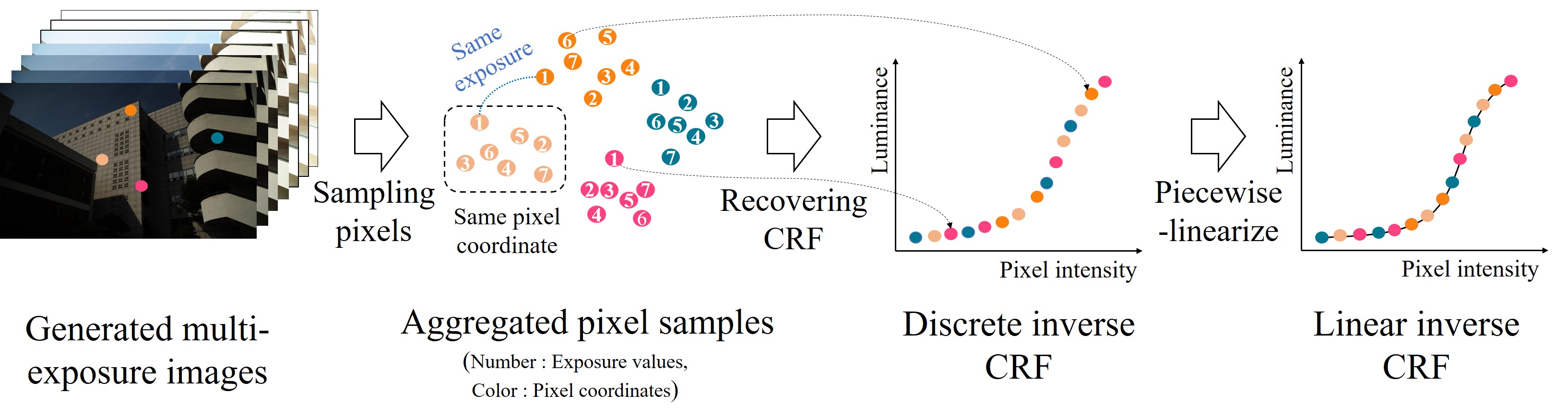}
  \caption{Conceptual diagram of the proposed piece-wise linearization for the CRF. We sample pixels from the multi-exposure stack to aggregate pixels of the same coordinate with different exposure values. We then estimate the inverse CRF with \cref{eq1} and convert the function into a differentiable linear form with the piece-wise linearization.}\label{fig_linearization}
\end{figure*}

\cref{fig_linearization} illustrates our approach to piecewise-linearize the inverse CRF. With the sampled pixels using the Grossberg and Nayar's method \cite{grossberg2003determining}, we linearize the function with the prior assumptions of the CRF having the characteristic of monotonically increasing with the shape of the non-linear curve. We reformulate the function with a piece-wise linear form to back-propagate the difference between the function value and the one before, as shown in \cref{eq3}. The simple linearization method enables the propagation of gradients to each pixel of the multi-exposure stack with the chain rule \cite{goodfellow2016deep}. The gradients from the loss of luminance values flow to pixel intensity values of each image, which imposes constraints on the generated multi-exposure stack to have correlated values with \cref{eq3}. Hence, our novel framework enables the networks to accomplish both the multi-exposure stack generation task and the HDR synthesis task, with the optimal objective of reconstructing high-quality HDR images. 

Furthermore, we implemented the polynomial curve fitting approach proposed by Mitsunaga and Nayar\cite{mitsunaga1999radiometric} in the differentiable HDR synthesis layer. Polynomial curves can be differentiated; however, as modern cameras' CRF have monotonic and resembling shape \cite{chen2019analyzing}, the higher-order models are not necessary. Therefore, we focused on the piece-wise linearization approach with further experiments (See the comparison between the piece-wise linearization and the polynomial curve fitting approach in the supplementary material). We verified our differentiable HDR layer outputs to reproduce the identical results with the MATLAB HDR Toolbox \cite{Banterle:2017} and to generate a pixel-wise gradient. 

\subsection{Recursive multi-exposure stack generation} We incorporate the recursive generation of the multi-exposure image stack with the prior knowledge of the exposure manifold space \cite{lee2018deepb}. We propose the recurrent-up and recurrent-down networks to be distinct from conventional methods \cite{lee2018deepa, lee2018deepb}. Since the process is defined as a recursive process, we implement the convolutional gated recurrent unit (Conv-GRU) \cite{siam2017convolutional} to construct the recurrent network. In addition, as multi-exposure images have different over-exposed and under-exposed regions regarding their exposure values, we decompose the exposure transfer task into two path-ways. From a given single image, our model learns the global tone and local details with the global network and local network, respectively. With decomposed images, the refinement network integrates global and local components to generate fine-tuned images. 

\begin{figure}
  \centering
  \includegraphics[width=0.87\columnwidth]{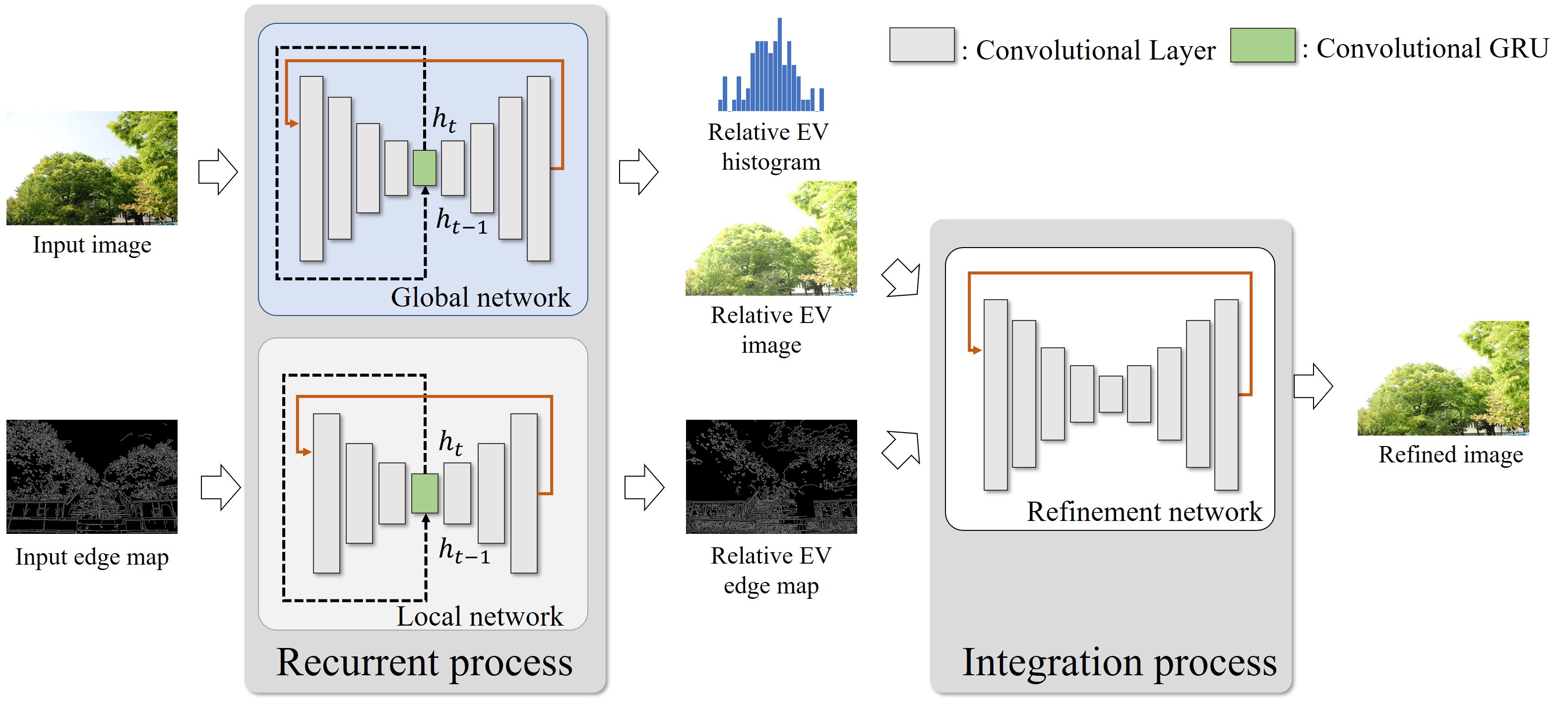}
  \caption{Sub-networks architecture. The global network focuses on minimizing the difference of histograms between the generated and target EV image, and the local network focuses on generating gradient-based edge structures. We facilitate the hidden state $h_t$ of $t$-the recursion to feed into the bottleneck layers of the global and local networks for the recurrent process. We then concatenate the input image, relative EV image, and edge map to feed into the refinement network to focus on the integration process.}\label{fig_subnetwork}
\end{figure}

\cref{fig_subnetwork} shows the structures of sub-networks in our model. Our recurrent-up and recurrent-down networks contain three sub-networks of U-Net structures \cite{ronneberger2015u} to transfer exposures to the images with the relative up and down EVs: the global, local, and refinement networks. The global and local networks are constructed with $5$-level and $4$-level structures, respectively, with $2$ convolutional layers for each level. We implemented the Swish activation \cite{ramachandran2017searching} on each convolution layer to alleviate a gradient vanishing problem in recurrent models. Note that the refinement network shares the same structure with the global network except for the Conv-GRUs on bottleneck layers. We impose the global and local networks to focus on adaptively responding to the number of recursions, and the refinement network to focus on integrating the global and local components, which are global tones and gradient-based edge structures of a target LDR image, respectively. The image decomposition approach resolves the complexity of a single network \cite{lee2018deepb}, as the adaptive response and integration are learned separately for the network to explicitly learn each corresponding task (See supplementary material for details).

The recurrent-up (or recurrent-down) network exploits the same weights for transferring exposures, even with the recurrent state that differs from the exposure value of an input. However, both the recurrent-up and recurrent-down networks should adaptively produce the over-exposed and under-exposed images corresponding to the exposure value of an input. Therefore, we use the conditional instance normalization to standardize feature maps of different exposure values. The normalization transforms a feature map, $X$, of which the shape is $C \times H \times W$, into a normalized map $Y$ by using two learnable parameters of $\gamma_{e}$ and $\beta_{e}$ with the target exposure value of $e$, which are in $\mathbb{R}^{C}$. The normalized map is formulated as $Y=\frac{\gamma_e(X-\mu)+\beta_e}{\sigma}$, where $\mu$ and $\sigma$ are the mean and the standard deviation of $X$ taken across spatial axes, respectively. In other words, our networks select the scale and shift factors according to the exposure value of an input LDR image. By using the conditional instance normalization, we can assist the network to focus on detecting subtle differences between the estimated and target images. Thus, we implemented a conditional instance normalization layer on the decoding layers of each level.

\subsection{Training} Our model is designed to facilitate the recurrent structure, which shifts the exposure level of the image gradually, as presented in \cref{fig_model}. Specifically, the recurrent-up and recurrent-down networks are trained separately with a given single LDR image to generate the multi-exposure stack recursively. For the sub-networks, the global and local networks are trained in advance for 10k iterations, then we jointly trained the entire network, including the refinement network. The loss functions are defined independently with each sub-network. Specifically, the global network is trained with the pixel-wise $L_1$ loss ($L_1$) and histogram loss ($L_{hist}$) to constraint the network to generate the image with a similar global tone to the target image (See the supplementary material for details). The local network is trained with pixel-wise $L_1$ loss ($L_{edge}$) on edge maps computed with Canny edge detector \cite{canny1986computational} of $\sigma=2$. The refinement network is trained with $L_1$ loss ($L_1$), the contextual bilateral loss ($L_{CoBi}$) \cite{zhang2019zoom}, and the HDR loss ($L_{HDR}$). For the HDR loss, we used a tone-mapped HDR loss with $\mu$-law to stabilize the training process \cite{yan2019attention}. Note that $L_{CoBi}$ alleviates the ghosting artifacts due to the misaligned images by minimizing the distances between the matching features extracted from the $3$-rd and $4$-th layer of the pre-trained VGG-19 network \cite{simonyan2014very} with the bilateral filtering. Overall loss functions are formulated as follows:
\begin{equation} \label{eq_loss1}
\begin{split}
&L_{global} =  \lambda_{1} L_{1} + \lambda_{2} L_{hist} \\
&= \frac{\lambda_{1}}{N \cdot E }\sum^{E}_{e}\sum^{N}_{i}|\hat{I}^{e}_{i} - I^{e}_{i}|\hspace{18mm}\\
&+ \frac{\lambda_{2}}{L\cdot E}\sum^{E}_{e}\sum^{L}_{l}|cnt_l(\hat{I}^e) - cnt_l(I^e)|,
\end{split}
\end{equation}
\begin{equation} \label{eq_loss2}
L_{local} = \lambda_{3}L_{edge} = \frac{\lambda_{3}}{N \cdot E}\sum^{E}_{e}\sum^{N}_i|\hat{E}^{e}_{i} - edge(I^{e}_{i})|,
\end{equation}
\begin{equation} \label{eq_loss3}
\begin{split}
&L_{refine} = \lambda_{4}L_{1} + \lambda_{5}L_{HDR} + \lambda_{6}L_{CoBi} \\
&= \frac{\lambda_{4}}{N \cdot E}\sum^{E}_{e}\sum_{i}^{N}|\hat{I}^{e}_{i} - I^{e}_{i}| +  \frac{\lambda_{5}}{N}\sum_{i}^{N}|log\frac{1+\mu\hat{H_i}}{1+\mu H_i}|\\ 
&+ \frac{\lambda_{6}}{M}\sum_j^M \min_{k} (\mathbb{D}_{p_j,q_k} + w_s \mathbb{D'}_{p_j,q_k}),
\end{split}
\end{equation}
where $N$, $E$, $L$, and $M$ denote the number of pixels, exposure values, intensity levels, and features respectively, and for all the equations, $\hat{\cdot}$ represents the prediction of the network. $I_{i}^{e}$ denotes the $i$-th pixel value in image $I$ of exposure value $e$, and $cnt_{l}(\cdot)$ indicates the number of pixels which has a rounded down intensity $l$ in the input image $I$. $edge(\cdot)$ extracts gradient-based edge maps from the image $I$, and $E_i$ denotes the $i$-th pixel value in predicted edge map. $H_i$ is a pixel luminance in the HDR image, and $\mu$ is the compression parameter of the HDR image, where we set the value with $5000$. $\mathbb{D}_{p,q}$ indicates the sum of cosine distances between all the matched features of $p$ and $q$, and ${\mathbb{D'}}_{p,q}$ indicates spatial coordinate distance. Note that $j$ and $k$ indicate indices of the matched feature of $p$ and $q$ respectively. We set the hyperparameters $\lambda_1 = \lambda_3 = \lambda_4 = \lambda_5 =1$ and $\lambda_2 = \lambda_6 = 0.1$ in our experiments to stably train the networks.

\begin{figure*} [ht]
  \centering
  \includegraphics[width=0.85\textwidth]{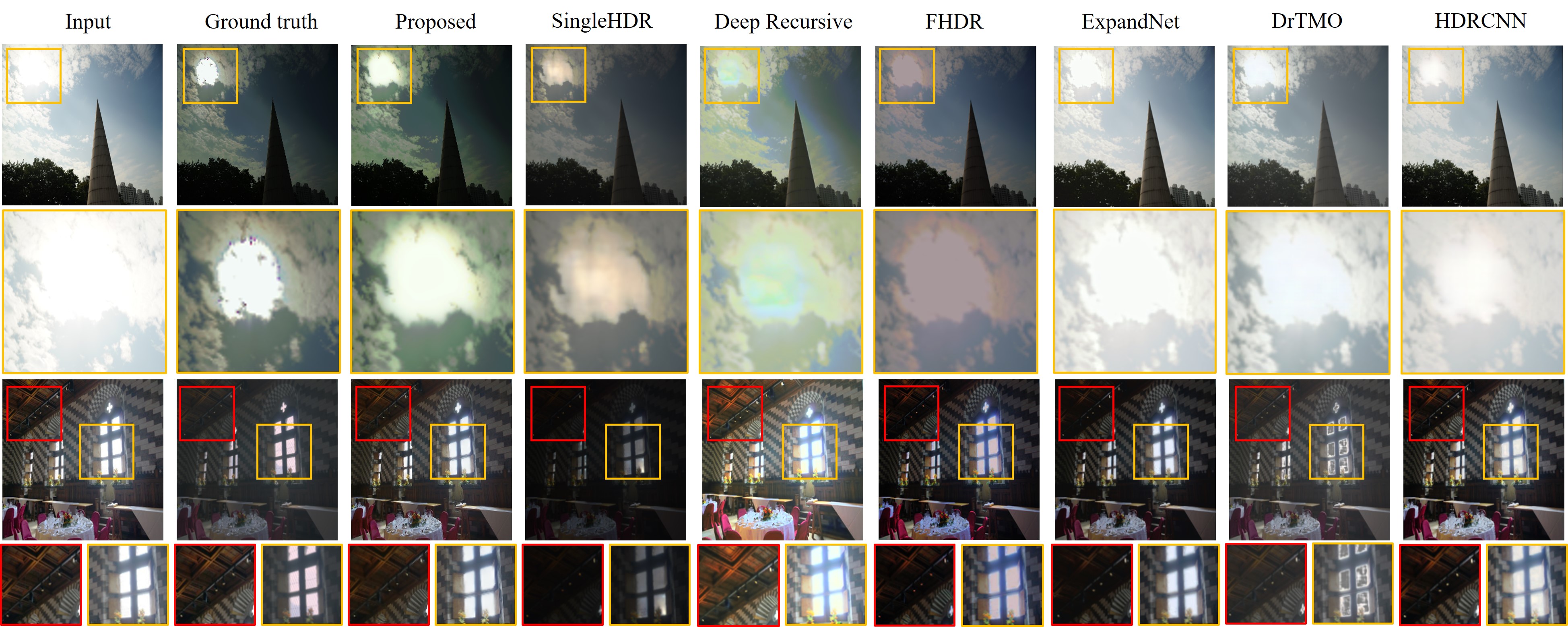}
  \caption{Comparison of tone-mapped HDR images from 6 different HDR reconstruction approaches on VDS, HDR-Eye, and RAISE datasets. The loss of image details in over-exposed and under-exposed regions occurs with the SingleHDR, FHDR, ExpandNet, and HDRCNN. The DrTMO and Deep recursive HDRI, which are stack-based methods, suffer from the local inversion artifacts. Nonetheless, our method reduces local inversion artifacts and preserves image details and contrasts in overexposed regions.} \label{fig_sota}
\end{figure*}

\begin{table*}
  \caption{Quantitative comparison of proposed and conventional HDR reconstruction methods. We measured the HDR-VDP-2 score \cite{mantiuk2011hdr} for synthesized HDR images.}\label{HDRVDP_table}
  \centering
  \begin{tabular}{l|l|c|c|c}
  \toprule
    \multirow{2}{*}{Method} &\multirow{2}{*}{Training dataset quantity} &VDS  &HDR-Eye &RAISE\\ \cline{3-5}
    & &$m\pm\sigma$ &$m\pm\sigma$ &$m\pm\sigma$ \\ \midrule
    \textbf{Proposed} & \textbf{48 scenes} & \textbf{58.807$\pm$5.413} & \textbf{55.914$\pm$1.917} &\textbf{59.493$\pm$3.420}\\ \hline
    HDRCNN  & 3,700 scenes &53.031$\pm$4.957 & 50.804$\pm$5.790 & 57.154$\pm$3.642 \\ \hline
    DrTMO  & 1,043 scenes & 55.227$\pm$4.662  &51.800$\pm$5.933 & 57.645$\pm$4.028\\ \hline
    Deep recursive HDRI  & \textbf{48 scenes} &56.347$\pm$3.492 & 52.832$\pm$2.944 & 57.570$\pm$3.697\\ \hline
    ExpandNet  & 1,013 scenes & 44.720$\pm$9.432 & 50.428$\pm$4.493 & 54.717$\pm$ 1.998\\ \hline
    FHDR & 39,460 scenes & 57.708$\pm$6.373 & 53.815$\pm$3.603 & 59.144$\pm$2.764\\
    \hline
    SingleHDR & 10,289 scenes & 55.237$\pm$4.487 & 54.509$\pm$3.714 & 59.304$\pm$ 3.541\\  
    \bottomrule
  \end{tabular}
\end{table*}

\section{Experimental Results}\label{gen_inst}
\subsubsection{Datasets} We trained our model on the VDS dataset \cite{lee2018deepa}, where the training set has 48 multi-exposure stacks, and the testing set has 48 stacks. In addition, we evaluated our model on the stacks of the HDR-Eye dataset \cite{lee2018deepa, liu2020single, nemoto2015visual}, which is widely used for the performance evaluation. To perform evaluations on more real image dataset, we conducted experiments with the RAISE dataset \cite{dang2015raise}. Input images were upscaled or downscaled into 256 $\times$ 256 pixel resolutions by the Lanczos interpolation method \cite{ken1990filters}, and all LDR images were in the sRGB color space.
\subsubsection{Implementation} For training the recurrent-up and recurrent-down networks, we chose the gradient centralized Adam optimizer \cite{yong2020gradient} with the learning rate of $1e^{-4}$. The momentum parameters of $\beta_1$ and $\beta_2$ were set to 0.5 and 0.999, respectively. We trained our model with a batch size of 1. Our model was trained on two GTX Titan X GPUs for four days to reach 80k iterations. 
\subsubsection{Evaluation metrics} We evaluated the quality of HDR image reconstruction with the HDR-VDP-2 score \cite{liu2020single, mantiuk2011hdr, marnerides2018expandnet}. The experiments were conducted under the same process provided with the state-of-the-art method \cite{marnerides2018expandnet, liu2020single}. We scaled the target and generated HDR image to match the 0.1 and 99.9 percentiles before measuring the HDR-VDP-2 score. We have set the hyperparameters of HDR-VDP-2 score as the color encoding of RGB-BT.709 and 30 pixels per one visual degree. We also assessed the quality of estimated multi-exposure stacks with peak signal-to-noise ratios (PSNR), structure similarity (SSIM), and multi-scale SSIM (MS-SSIM). We then used the Reinhard tone mapping operator \cite{reinhard2002photographic} for the visualization.

\subsection{Comparison with the state-of-the-art methods}
The comparison evaluations were performed with 6 recent deep learning-based methods, both direct methods(HDRCNN \cite{eilertsen2017hdr}, ExpandNet \cite{marnerides2018expandnet}, FHDR \cite{khan2019fhdr}, SingleHDR \cite{liu2020single}) and multi-exposure stack-based methods (DrTMO \cite{endo2017deep}, Deep recursive HDRI \cite{lee2018deepb}) as benchmarks. The interchangeability of training datasets between methods is limited as the direct methods need a large amount of LDR-HDR image pair datasets, and the multi-exposure stack-based method requires an adequate amount of images of different exposures. Therefore, we used pre-trained models for ExpandNet, DrTMO, FHDR (2-iteration), and SingleHDR.

\subsubsection{HDR quality assessment} When measuring the quality of generated HDR images, we applied the method of Debevec and Malik \cite{debevec2008recovering} with the stack-based methods. The size of training datasets across different methods was imbalanced, as shown in \cref{HDRVDP_table}. Compared to other models, our method is trained with much fewer scenes and outperformed both the direct and multi-exposure stack-based methods with favorable HDR-VDP-2 scores on three datasets. The result indicates that our method has a strong advantage in the data efficiency. 

\begin{figure}
\centering
\includegraphics[width=0.95\columnwidth]{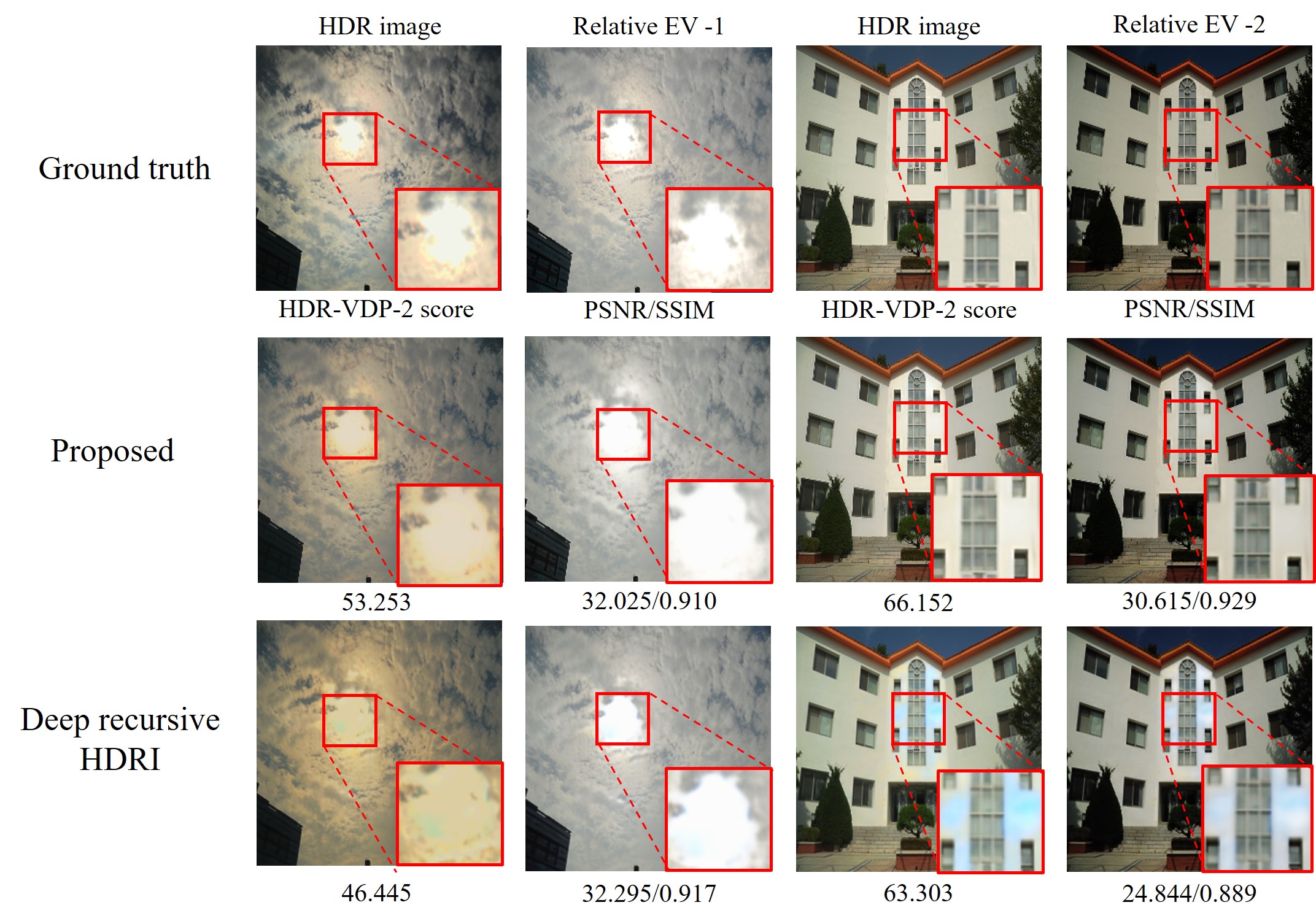}
\captionof{figure}{Case analysis of correlations between the multi-exposure stack reconstruction and the HDR reconstruction on the VDS dataset. The experiment was conducted with Lee \etal \cite{lee2018deepb} and our method. The result shows that two factors (stack reconstruction accuracy, HDR reconstruction accuracy) have a weak correlation (suboptimal, optimal).} \label{fig_stack}
\end{figure}

\begin{table*}[t]
  \caption{Quantitative comparison of stack reconstruction results. Relative EV+1 indicates the average value of three recursive recurrent-up results and Relative EV-1 indicates the average value of three recurrent-down results.}\label{stack_table}
  \centering
  \begin{tabular}{c|c|c|c|c}
    \toprule
    \multirow{2}*{} & \multirow{2}*{Method}  &{PSNR (dB)} &{SSIM}  &{MS-SSIM} \\ \cline{3-5} 
    & & $m\pm\sigma$ & $m\pm\sigma$ & $m\pm\sigma$ \\ \midrule
    \multirow{2}*{\begin{tabular}[c]{@{}c@{}}Relative\\EV +1\end{tabular}} & \textbf{Proposed} & 30.292$\pm$3.725   & 0.952$\pm$0.050   & 0.989$\pm$0.009 \\\cline{2-5} 
     & Deep recursive HDRI \cite{lee2018deepb} & 30.142$\pm$2.873   &0.955$\pm$0.036 &0.986$\pm$0.010 \\
    \hline 
    \multirow{2}*{\begin{tabular}[c]{@{}c@{}}Relative\\EV -1\end{tabular}} & \textbf{Proposed} &30.403$\pm$3.601   & 0.940$\pm$0.038   & 0.985$\pm$0.011 \\\cline{2-5} 
    & Deep recursive HDRI \cite{lee2018deepb} & 30.483$\pm$3.836  & 0.936$\pm$0.044 &0.982$\pm$0.014 \\
   \bottomrule
  \end{tabular}
\end{table*}

\subsubsection{Multi-exposure stack reconstruction} We verified the relations between the multi-exposure stack reconstruction and the HDR reconstruction. Specifically, we evaluated PSNR, SSIM, and MS-SSIM results of reconstructed stacks by our method and the previous stack-based method \cite{lee2018deepb}. The previous approach \cite{lee2018deepb} focused on reconstructing the multi-exposure stack, and hence, reproducing stacks with high PSNRs, SSIMs, and MS-SSIMs. However, with the results of \cref{fig_stack} and \cref{stack_table}, our method reproduced similar PSNR, SSIM, and MS-SSIM with the previous method, but achieved much higher HDR-VDP-2 scores. The results indicate that focusing on the exposure transfer task might lead to suboptimal generation performances. Furthermore, our method does not include any adversarial loss; however, as the direct relation between pixel values was imposed during the training, we achieved the result of the highest quality, thereby providing higher HDR-VDP-2 scores.

\subsection{Ablation studies}
We evaluated the effectiveness of the individual components in our model on the VDS dataset, as shown in \cref{Ablation}. We added modules incrementally on the U-Net structure \cite{ronneberger2015u}, which is a baseline of our model with $5$-level and $2$ convolutional layers for each level, and evaluated with the HDR-VDP-2 score. The overall results show that our method using all modules improved $9.305$ and $4.483$ with HDR-VDP-2 score and PSNR, respectively. 
\subsubsection{Recurrent network} First, we added the recurrent module, the Conv-GRU \cite{siam2017convolutional}, to be located in the bottleneck layer. We utilized the hidden state of each recurrent network to convey the important state variables, such as recursion numbers to the network. \cref{Ablation} shows that recurrent module could increase both the HDR reconstruction performance with the HDR-VDP-2 score and multi-exposure stack reconstruction with PSNR by $2.842$ and $1.788$, respectively. 
\subsubsection{Conditional instance normalization} We demonstrated the effectiveness of the conditional instance normalization layer with a comparison experiment with the instance normalization layer \cite{ulyanov2016instance}.  We confirmed that the conditional instance normalization layer decreases the standard deviation of the reconstruction error.
\subsubsection{Image decomposition} We decomposed input images into global and local components. To verify the effectiveness of our structure, we compared the PSNR result of the decomposition network with that of the baseline network, as shown in \cref{Ablation}. We trained both networks for the same iterations, and the quantitative result of PSNR shows that decomposition decreases the reconstruction error. 
\subsubsection{Differentiable HDR synthesis layer} The proposed differentiable HDR synthesis layer could reconstruct the target HDR image without any learnable parameters in the layer. The mean of HDR-VDP-2 score was significantly increased by up to $3.265$, and the standard deviation was decreased by up to $1.270$. Hence, the differentiable HDR synthesis layer guided the network to generate the high-quality HDR image while stabilizing the training process.
\subsubsection{Contextual bilateral loss} To enhance the perceptual quality of the generated multi-exposure stack, we added contextual bilateral loss \cite{zhang2019zoom} to fine-tune our networks. This loss alleviated the limitations of using ghosting artifacts induced by applying $L_1$ loss on the misaligned image dataset. \cref{Ablation} shows that contextual bilateral loss fine-tunes the outputs of networks.

\begin{table}
  \caption{Performance of various configurations on the VDS dataset \cite{lee2018deepa}}\label{Ablation}
  \centering
  \begin{tabular}{l|c|c}
  \toprule
    Method & HDR-VDP-2 & PSNR (dB)  \\ \midrule
    Baseline  &\small{49.502$\pm$6.519}  & \small{25.864$\pm$3.013}\\ \hline
     + Recurrent network & \small{52.344$\pm$6.852}  &\small{27.652$\pm$3.189}\\ \hline
    \begin{tabular}[l]{@{}l@{}} + Conditional instance\\\hspace{2mm}normalization\end{tabular} & \small{53.020$\pm$5.110} &\small{27.996$\pm$2.779}\\ \hline
     + Image decomposition &\small{54.548$\pm$6.455}  & \small{28.542$\pm$3.500}  \\ \hline
   \begin{tabular}[l]{@{}l@{}}  + Differentiable HDR\\\hspace{2mm}synthesis layer\end{tabular}
   &\small{57.813$\pm$5.185}  &\small{29.592$\pm$3.596}   \\\hline
   \begin{tabular}[l]{@{}l@{}}  + Contextual bilateral\\\hspace{2mm}loss\end{tabular}
    &\small{58.807$\pm$5.413}  & \small{30.347$\pm$3.663}  \\
  \bottomrule
  \end{tabular}
\end{table}

\section{Conclusion}
This paper presented a novel framework that generates both the multi-exposure stack and the HDR image. We proposed a differentiable HDR synthesis layer with deep learning framework that converts the HDR synthesis process to be differentiable with the linear approximation technique. Hence, our approach enabled an entire network to be trained to reconstruct HDR images with direct supervision. Moreover, we used recurrent and decomposition approaches for the multi-exposure stack generation with the purpose to disentangle the exposure transfer task. The results show that our framework achieved the state-of-the-art results for both direct and stack-based methods by removing the severe local inversion artifacts and restoring the details regardless of image conditions. For the future work, as we yielded impressive results regarding the relatively low PSNR, we will further analyze the relationship between the multi-exposure stack generation and the HDR image synthesis to optimize multiple tasks to be mutually complementary. 

\section{Acknowledgements}
This research was supported by the MSIT (Ministry of Science and ICT), Korea, under the
ITRC (Information Technology Research Center) support program (IITP2020-2018-0-01421) supervised by the IITP (Institute for Information Communications Technology Planning Evaluation), and the National Research Foundation of Korea (NRF) grant funded by
the Korea government (MSIT) (No. 2018R1D1A1B07048421)

\section{Ethics Statement}
As we focus on theoretical grounds for restoring the HDR image, it has both positive and negative sides. On the negative side, our method is prone to be exploited like face generation and autonomous system attack \cite{muller2020ethics}, because our method restores lost information based on the context of an input image. However, realistic images can be easily acquired from LDR images taken by a standard camera through our method. Depending on the applications using this method (e.g., autonomous driving, etc.), users should consider foreseeable potential risks, where the contents of the output images may be changed by adversarial attacks for the recurrent-up or recurrent-down networks.

\bibliography{references}

\end{document}